\RequirePackage{fix-cm}
\documentclass[smallcondensed]{svjour3}     
\smartqed  
\usepackage{graphicx}
\usepackage{csquotes}
\usepackage{hyperref}
\usepackage{cite}
\usepackage{amsmath,amssymb,amsfonts}
\usepackage{algorithmic}
\usepackage{textcomp}
\usepackage{framed,multirow}
\usepackage{float}
\usepackage{mathrsfs}
\usepackage{footnote}

\begin{document}
	
	\title{CHS-Net: A Deep learning approach for hierarchical segmentation of COVID-19 via CT images}
	
	\titlerunning{CHS-Net}        
	
	\author{Narinder Singh Punn*         \and
		Sonali Agarwal 
	}
	
	
	\institute{Narinder Singh Punn* \at
		IIIT Allahabad, Prayagraj, India, 211015 \\
		Tel.: +91-7018466740\\
		\email{pse2017002@iiita.ac.in}           
		\and
		Sonali Agarwal \at
		IIIT Allahabad, Prayagraj, India, 211015 \\
		Tel.: +91-9415647042\\
		\email{sonali@iiita.ac.in}
	}
	

	\maketitle
	
	\begin{abstract}
		The pandemic of novel severe acute respiratory syndrome coronavirus 2 (SARS-CoV-2) also known as COVID-19 has been spreading worldwide, causing rampant loss of lives. Medical imaging such as computed tomography (CT), X-ray, etc., plays a significant role in diagnosing the patients by presenting the visual representation of the functioning of the organs. However, for any radiologist analyzing such scans is a tedious and time-consuming task. The emerging deep learning technologies have displayed its strength in analyzing such scans to aid in the faster diagnosis of the diseases and viruses such as COVID-19. In the present article, an automated deep learning based model, COVID-19 hierarchical segmentation network (CHS-Net) is proposed that functions as a semantic hierarchical segmenter to identify the COVID-19 infected regions from lungs contour via CT medical imaging using two cascaded residual attention inception U-Net (RAIU-Net) models. RAIU-Net comprises of a residual inception U-Net model with spectral spatial and depth attention network (SSD) that is developed with the contraction and expansion phases of depthwise separable convolutions and hybrid pooling (max and spectral pooling) to efficiently encode and decode the semantic and varying resolution information. The CHS-Net is trained with the segmentation loss function that is the defined as the average of binary cross entropy loss and dice loss to penalize false negative and false positive predictions. The approach is compared with the recently proposed approaches and evaluated using the standard metrics like accuracy, precision, specificity, recall, dice coefficient and Jaccard similarity along with the visualized interpretation of the model prediction with GradCam++ and uncertainty maps. With extensive trials, it is observed that the proposed approach outperformed the recently proposed approaches and effectively segments the COVID-19 infected regions in the lungs. 
		
	\end{abstract}
	
	\section{Introduction}
\begin{figure}[H]
		\centering
		\includegraphics[width=\columnwidth] {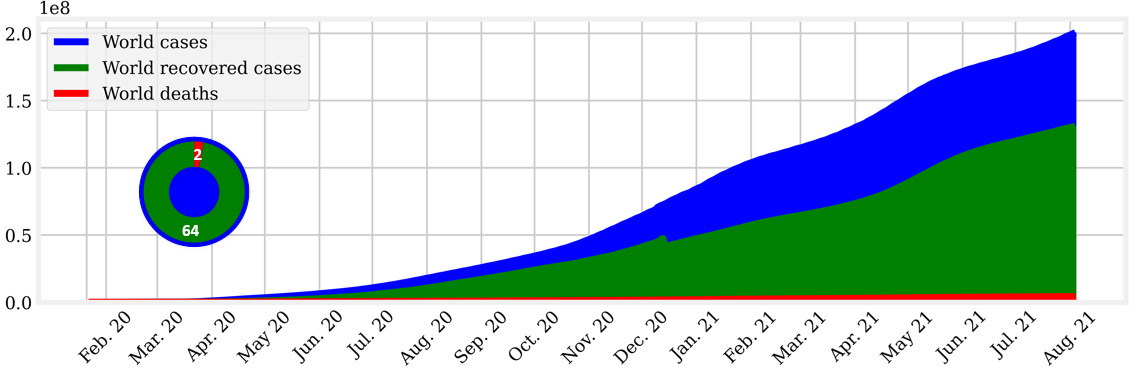}
		\caption{Exponential growth trend of COVID-19 worldwide.}
		\label{fig1}
	\end{figure}

	The novel coronavirus, also known as COVID-19, is an on-going worldwide pandemic that initiated from Wuhan, the People's Republic of China in December 2019 and till August 4, 2021, have caused 200,719,425 infections and 4,256,660 deaths worldwide~\cite{WHO2}. The exponential growing trend of COVID-19 is highlighted in Fig.~\ref{fig1} that shows the number of confirmed cases out of which 2\% died and 64\% recovered worldwide since the time it is recorded~\cite{jhcgithub}. With the exponential spread of the virus, the World Health Organization (WHO) declared the coronavirus outbreak as a public health emergency of international concern (PHEIC) in January 2020 and later as a pandemic in March 2020~\cite{WHO1}. This has raised concern in every sector of the international community such as public health, transportation, marketing, tourism, manufacturing, lifestyle, etc. Even with various advancements in technology, unfortunately, till now there is no concrete solution or medicine to cure COVID-19 and hence the international community is adopting the avoidance and preventive measures that involve self hygiene, no social contact, avoiding finger touch on the public doors, elevators, etc.~\cite{punn2020monitoring}. Since COVID-19 is highly contagious, the infected ones are kept in isolation and closely monitored by doctors and experts for treatment to minimize its spread. Moreover, the availability of resources of COVID-19 detection and diagnosis is quite limited as compared to its requirement, hence researchers are exploring all possible ways to detect and analyse the impact of infection on the human body. With this motivation, biomedical image analysis has become a prominent area of research to aid in the diagnosis of COVID-19.
	
	In biomedical image analysis, the problems can be interpreted as classification and segmentation to identify and detect any abnormality in the radiography~\cite{rajinikanth2020harmony} via deep learning techniques, where the convolution neural network (CNN) based architectures are the most promising and popular choice in the research community. CNNs have displayed remarkable performance over the years and are being deployed for endoscopic videos~\cite{gomez2019low}, CT images~\cite{choe2019deep}, diagnosis of pediatric pneumonia using chest X-ray images~\cite{punn2020automated, kermany2018identifying}, etc. In the context of COVID-19, the classification task involves a prediction for the patient being infected with the virus in the presence of binary or multi-class samples~\cite{punn2020automated, li2020ct, singh2020classification} (involving other viruses or diseases than COVID-19), whereas in segmentation the coronavirus infected regions are localized and in-painted~\cite{fan2020inf, shan2020lung} via lungs CT or X-ray imaging.
	
	\subsection{Why lungs segmentation?}
\begin{figure}
		\centering
		\includegraphics[scale=0.4] {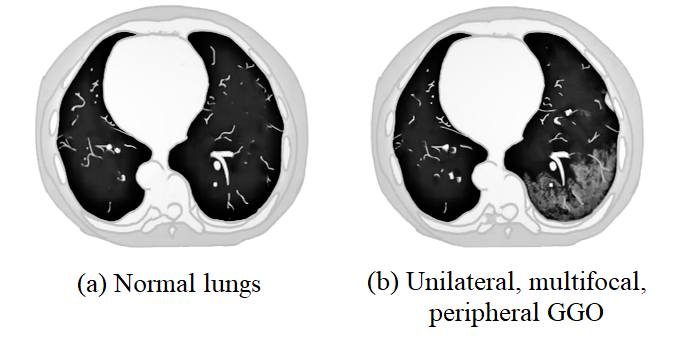}
		\caption{Chest CT imaging of normal and COVID-19 infected lungs.}
		\label{fig2}
	\end{figure}

	Currently, reverse transcription-polymerase chain reaction (RT-PCR) test acts as the standard to diagnose and confirm the symptoms of COVID-19 in any patient~\cite{ai2020correlation}. However, the RT-PCR assay is deficit to fulfil the demand in every area. The test takes around 4 to 6 hours and is less sensitive to confirm the coronavirus at the initial stages. The current findings indicate that COVID-19 affects various organs of a human being, such as blood vessels, heart, stomach, intestines, brain and kidneys~\cite{wadman2020does}. The virus enters into the cells surface receptors angiotensin-converting enzyme 2 or ACE2 which is present on alveoli of the lungs. Therefore, lungs become the primary target for the virus affection which later spread to other body organs. Following this context, computed tomography (CT) imaging of the human lungs is considered to diagnose and test COVID-19 infections. It has been observed that bilateral, multifocal and peripheral ground glass opacification (GGO) that follows typical patterns, are predominant CT findings in patients suffering from COVID-19~\cite{ChestCT} as highlighted in Fig.~\ref{fig2}. However, for any radiologist analyzing CT scan is a time consuming and tedious task. Thus, biomedical image analysis techniques involving deep learning and machine learning algorithms are developed for faster cures and treatment.
	
	\begin{figure}[]
		\centering
		\includegraphics[scale=0.45] {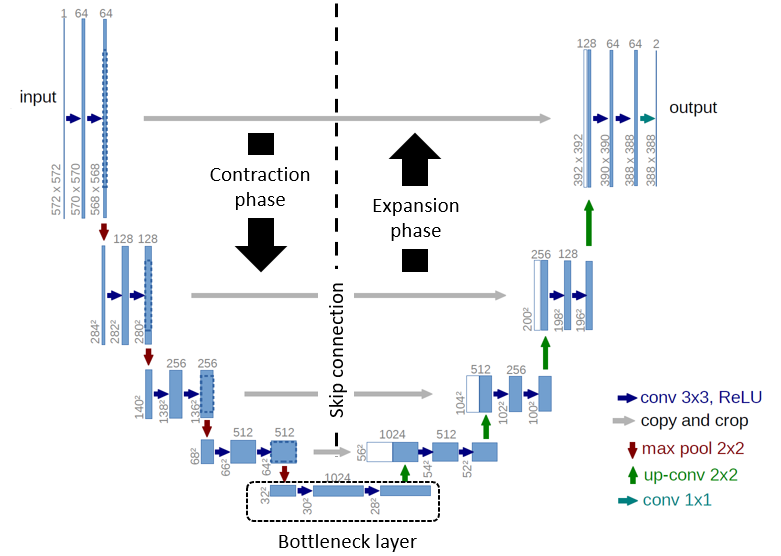}
		\caption{Schematic representation of base U-Net model.}
		\label{fig3}
	\end{figure}
	
	\subsection{U-Net}
	U-Net~\cite{ronneberger2015u} model, as shown in Fig.~\ref{fig3}, is the most extensively utilized CNN based deep learning architecture for medical image segmentation. Due to its symmetrical encoder-decoder framework divided into contraction and expansion paths, the model can extract low and high-level features at the varying hierarchy of resolutions, and reconstruct the output segmentation map in the desired dimensions. 
	
\begin{figure}[]
		\centering
		\includegraphics[width=\columnwidth] {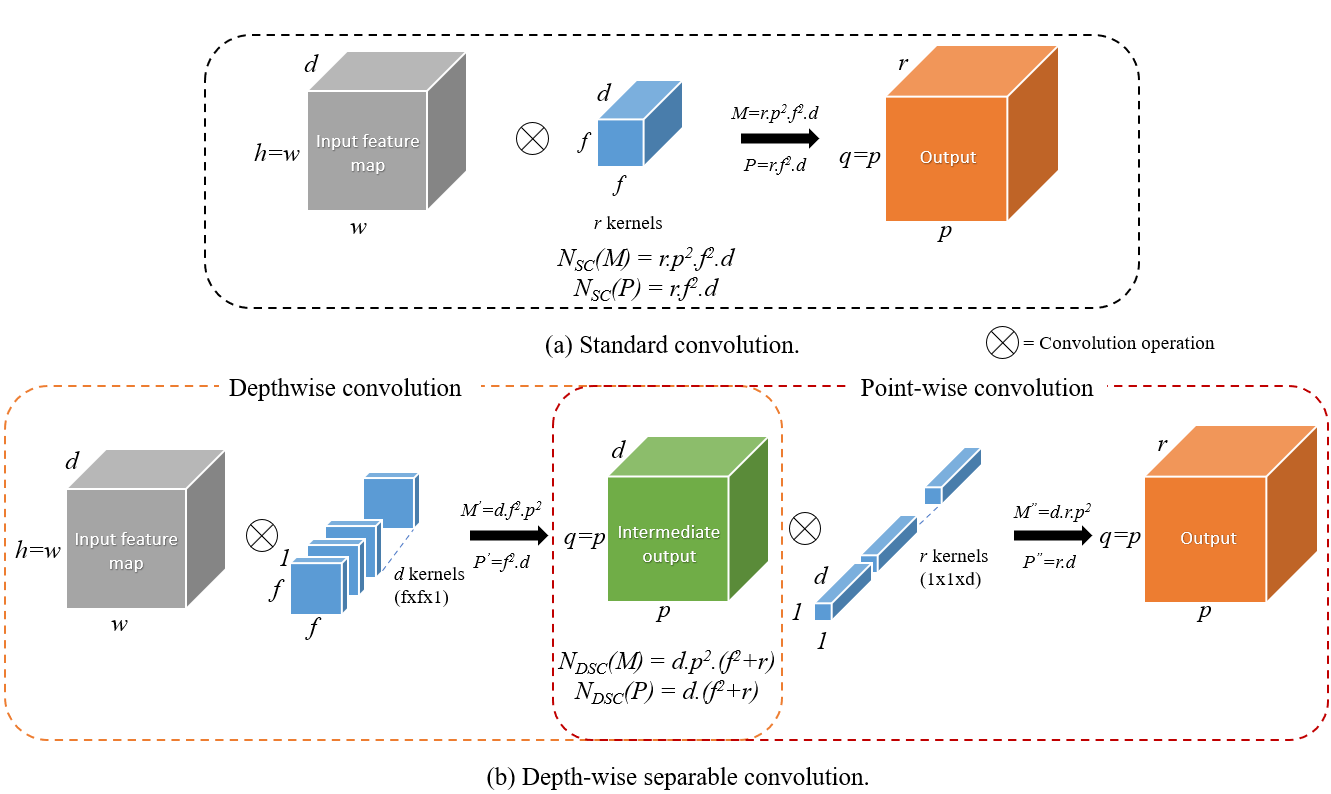}
		\caption{Standard convolution operation vs depthwise separable convolution operation.}
		\label{fig4}
	\end{figure}

	Following this, many variations of U-Net have been proposed~\cite{punn2021modality}. In most of the extended versions of U-Net, the feature maps produced in the contraction phase are preprocessed via attention gates (AG)~\cite{oktay2018attention}, squeeze and excitation block (SE)~\cite{hu2018squeeze}, spatial and channel SE blocks~\cite{roy2018concurrent}, etc., before concatenating with the corresponding expansion layer. It has also been observed that to segment regions of varying shapes and size require different sizes of receptive field~\cite{punn2020inception}. Since, the COVID-19 infected regions may vary in shape, size and location, the present article incorporates the proposed inception block into the standard U-Net. Furthermore, to improve the computational power,  the proposed approach also integrates depthwise separable convolution (DSC) layers~\cite{chollet2017xception}, divided into two stages: depthwise and pointwise convolutions. Fig.~\ref{fig4} draws the contrast between standard convolution (SC) and DSC for some feature map,  $\mathcal{F}\in\mathbb{R}^{w\times h\times d}$, where $w$ is width, $h$ is height and $d$ is depth of the input. It is observed that DSC reduces the number of multiplications (M) and parameters (P) than the SC as given by the Eq.~\ref{eq13}. This significantly reduces the training time and computational cost without affecting the performance of the model.
	
	\begin{equation}
		\frac{N_{DSC}}{N_{SC}}=\frac{1}{r}+\frac{1}{f^{2}}
		\label{eq13}
	\end{equation}
	where $N_{DSC}$ and $N_{SC}$ indicates number of parameters or multiplications in depthwise separable convolution and standard convolution respectively, $r$ is the depth of the output volume and $f$ indicates dimension of the kernel as shown in Fig.~\ref{fig4}.
	
	\subsection{Why hierarchical segmentation?}
	The present article addresses the challenging problem of efficiently identifying the COVID-19 infected regions in the CT images. Since these infected regions are present inside the lungs, the information present outside the lungs area becomes irrelevant. Therefore, in the proposed approach instead of direct segmentation, a hierarchical segmentation approach is introduced. In hierarchical segmentation, two residual attention inception U-Net (RAIU-Net) models are cascaded where the first model extracts lungs region from the CT images to generate the lungs contour feature maps and the second model utilizes these maps to segment the COVID-19 infected areas. 
	
	\subsection{Challenges addressed}
	COVID-19 infection segmentation in CT images is a challenging task due to the following concerns:
	\begin{enumerate}
		\item The presence of high variation in pattern, area and locale of infections in CT slices makes it difficult to segment. For instance, small infected regions can easily get neglected by the model which increases the false-negative predictions. This challenge is addressed by proposing inception convolution blocks that follow depthwise separable convolutions (DSC) of varying filter sizes ($1\times 1$, $3\times 3$, $5\times 5$) and a hybrid pooling layer accompanied with batch normalization and rectified linear unit activation.
		\item Limited data availability of the COVID-19 infected patients, resulting due to privacy and security concerns. This affects the training of deep learning models. This challenge is addressed by the segmentation loss function and fusion of publicly accessible multiple datasets of CT volumes consisting of coronavirus and non-coronavirus slices to generate a large volume of data. 
		\item It is observed that the intensity variance between the infected regions and background (regions outside the lungs area) is small, this restricts the deep learning models to identify the infected regions efficiently. This challenge is tackled by proposing a hierarchical segmentation approach where the irrelevant background is discarded before the identification of COVID-19 infection by generating lungs contour maps.
		\item The proposed CHS-Net model is a deep network, where the deep networks suffer from performance degradation due to the problem of vanishing and exploding gradient. To address this problem, each block of the RAIU-Net model is equipped with residual (skip) connections to improve the flow of information in the network.
	\end{enumerate}
	
	\subsection{Our contribution}
	This article presents the following contribution in the COVID-19 infectious image segmentation research:
	\begin{enumerate}
		\item A novel deep learning hierarchical approach, CHS-Net, built using RAIU-Net, is proposed for segregating the coronavirus infected areas using CT scans by exploiting the potential strategies of the state-of-the-art deep learning models. 
		\item A residual inception module is incorporated with a U-Net model to efficiently decode the semantic and varying resolution information. 
		\item A hybrid of max pooling and spectral pooling is proposed for the efficient reduction in the spatial dimension of the feature maps with minimal loss of information.
		\item A skip connection based on spectral spatial and depth attention (SSD) mechanism is proposed that uses global spectral-max pooling to infer the inter-spatial and channel features correlations for the effective flow of feature maps between the contraction and expansion phases.
		\item A fusion dataset is introduced with 3560 CT slices, developed using COVID-19 CT segmentation nr.2 dataset~\cite{datasetcovid19ct} and COVID-19 CT lung and infection segmentation dataset~\cite{datasetcovid19ct2}. Each CT slice has the corresponding lungs mask and COVID-19 infection mask. The code and dataset are available at the github repository.\footnote{\url{https://git.io/Jtec9}\label{chsnet}}.
	\end{enumerate}
	
	\subsection{Article organization}
	The rest of the paper is presented in various sections involving a literature survey in the related work section which highlights the recent findings and approaches for COVID-19 detection via CT medical imaging. The later section discusses the proposed approach to effectively identify and segregate the COVID-19 infected regions in the lungs. Furthermore, the experimental and results sections describe the obtained results along with the exhaustive experimental trials and comparative analysis, dataset description and ablation study. The final section highlights the concluding remarks and further possible extensions of the work.
	
	\section{Related work}
	With rapid advancements in technology, many artificial intelligence driven solutions are being developed to fight against COVID-19 pandemic~\cite{agarwal2020unleashing}. In recent studies~\cite{li2020coronavirus, ding2020chest, meng2020ct}, CT abnormalities corresponding to COVID-19 are being utilized by practitioners and doctors. It is observed that CT scan highlights discrete patterns to identify the infected patients even at the initial stages, making automatic CT medical imaging analysis a promising area of research among the research community~\cite{meng2020ct}. It is also observed that CT diagnosis for COVID-19 abnormality detection can be carried before the appearance of clinical symptoms~\cite{ChestCT}. Hence, many research works have been proposed for automatic early detection with classification and segmentation of the COVID-19 infection from CT scans~\cite{shi2020review, shoeibi2020automated}.
	
	Li et al.~\cite{li2020artificial} proposed a fully automatic CNN based COVID-19 detection neural network (COVNet), to classify COVID-19 abnormalities from community acquired pneumonia (CAP) and normal cases using chest CT imaging. The authors achieved 96\% area under the receiver operating characteristic curve (AUC-ROC) to identify COVID-19 cases and performed better than RT-PCR testing. However, the proposed approach is not effective in segregating and classifying different types of pneumonia due to the limited data availability. Butt et al.~\cite{butt2020deep} proposed deep learning based COVID-19 screening system to distinguish covid infected samples from non-covid samples. The proposed system yielded faster detection rate than RT-PCR testing with an overall accuracy score of 86.7\%. Shan et al.~\cite{shan2020lung} proposed V-Net~\cite{milletari2016v} based deep learning model to segment and quantify the COVID-19 infected regions. The authors achieved a dice similarity index of 91.6\% $\pm$ 10.0\% between manual and deep learning enabled automatic delineation. However, these approaches do not provide localized information about the infected regions in the CT scan of the lungs.
	
	A multistage deep learning framework is proposed by Gozes et al.~\cite{gozes2020rapid} that follows segmentation to remove the irrelevant regions and classification of segmented regions into coronavirus infected and other viral pneumonia. For segmentation, the U-Net~\cite{ronneberger2015u} model is utilized to acquire the relevant regions and then a pretrained ResNet-50~\cite{he2016deep} model is fine-tuned to classify COVID-19 infected samples. Yan et al.~\cite{yan2020covid} proposed COVID-SegNet accompanied with feature variation block and progressive atrous convolutions to highlight the diverse infected regions along with the boundaries. The proposed approach achieved a dice score of 0.726 for COVID-19 segmentation. Furthermore, Hu et al.~\cite{hu2020weakly} developed an object detection based approach to highlight the infected region with the help of the bounding boxes. The authors followed a weakly supervised approach to improve model performance with a limited number of labelled COVID-19 samples. The authors employed VGG model variants to classify COVID-19 from CAP and non-pneumonia cases. In another approach, Oulefki et al.~\cite{oulefki2021automatic} proposed an image enhancement based segmentation approach to efficiently highlight COVID-19 infected lung regions. While extracting lung regions each lung is segregated to which local contrast enhancement is applied, thereby providing more details about target regions for better segmentation results. Fan et al.~\cite{fan2020inf} proposed lung infection segmentation deep network (Inf-Net) to segment COVID-19 infected regions with ground glass opacities (GGO) and consolidation while also addressing the challenges of high variation characteristics and low intensities of the abnormalities, and limited availability of the infected samples. With extensive trials, Inf-Net outperformed the recently proposed approaches.
	
	Recently, Mu et al.~\cite{mu2021progressive} proposed an improved encoder-decoder architecture named progressive global perception and local polishing (PCPLP) network. To generate finer segmentation results, the network is equipped with an improved attention strategy and multi-scale multi-level feature recursive aggregation (mmFRA) module that learns global feature representations concerning infection regions. The authors achieved promising results with a dice value of 0.78 while outperforming other state-of-the-art models. In similar work, an evolvable adversarial learning strategy is proposed by He et al.~\cite{he2021evolvable}, where three different mutation operators are utilized to train the generator with added gradient penalty for producing stable COVID-19 infection segmentation. However, for real word implications of such approaches, it is critical to quantify the prediction uncertainty of the model~\cite{abdar2021review}. Following from these notions the present article contributes towards further improvement in the COVID-19 segmentation performance by introducing a hierarchical segmentation approach with cascaded residual attention inception U-Net models that generates lungs contour maps to efficiently segment the coronavirus infected regions. In addition, uncertainty maps are generated using the Monte Carlo dropout strategy.
	
	\section{Proposed network}
\begin{figure}[!b]
		\centering
		\includegraphics[width=\textwidth] {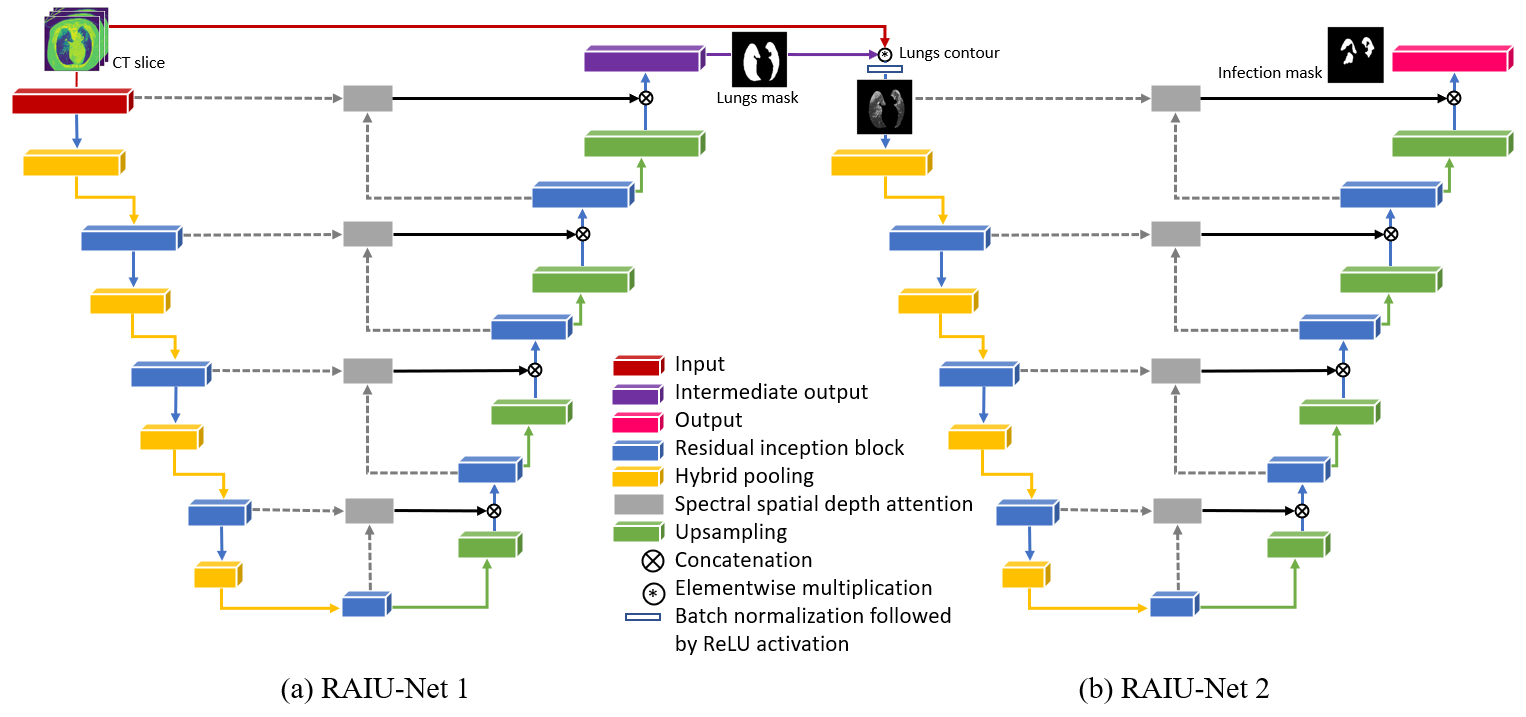}
		\caption{Schematic representation of CHS-Net (a) generates lungs contour maps, (b) generates infected regions.}
		\label{fig5}
	\end{figure}

	The proposed COVID-19 hierarchical segmentation network (CHS-Net), as shown in Fig.~\ref{fig5}, is inspired from the state-of-the-art deep learning architectures: U-Net~\cite{ronneberger2015u}, Google’s inception model~\cite{szegedy2015going}, residual network~\cite{he2016deep} and attention strategy ~\cite{oktay2018attention}. For hierarchical segmentation, two RAIU-Net models are connected in a series where the first model generates lungs contour maps (Fig.~\ref{fig5}(a)) and then the second model utilizes these maps to identify the infected regions (Fig.~\ref{fig5}(b)). The depth of the model is divided into four stages where each stage extract feature maps at different spatial dimensions. In contraction phase, each stage reduces the width and height by $50\%$, and increase the depth by $50\%$ and vice-versa in the expansion phase. 
	
\begin{figure}[]
		\centering
		\includegraphics[scale=0.35] {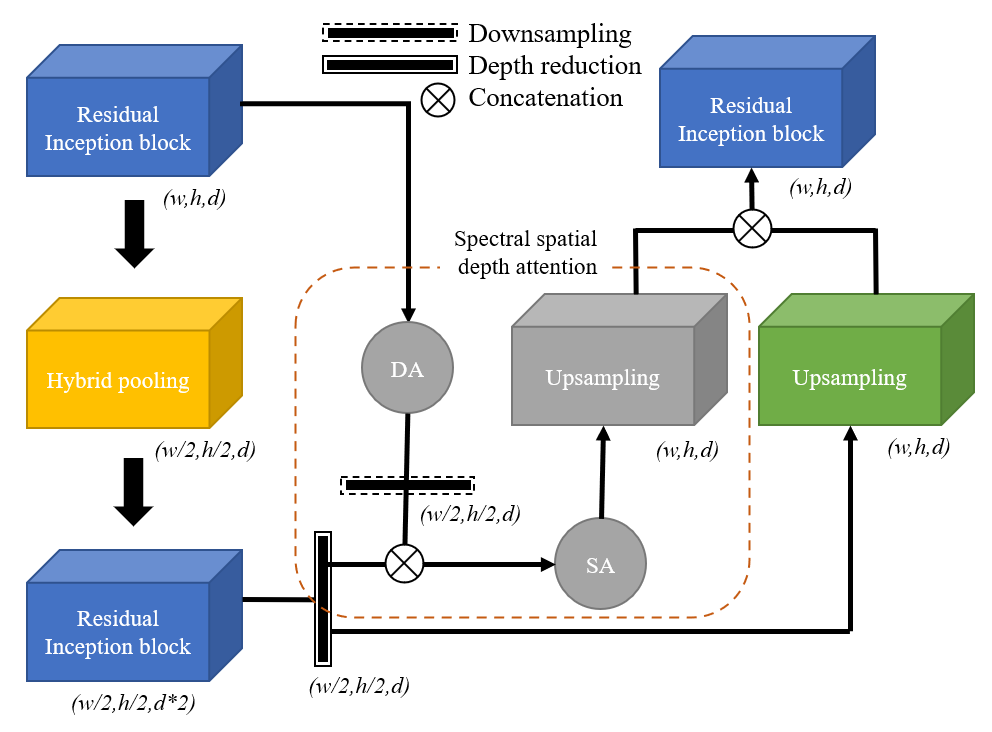}
		\caption{Building block of RAIU-Net.}
		\label{fig6}
	\end{figure}

	Fig.~\ref{fig6} presents the schematic representation of the building block of RAIU-Net for some input feature map,  $\mathcal{F}\in\mathbb{R}^{w\times h\times d}$. RAIU-Net model is developed in a U-Net fashion where each 2D convolution is replaced with inception blocks (concatenated 1$\times$1, 3$\times$3 and 5$\times$5 DSCs, and hybrid pooling followed by batch normalization to reduce the internal covariance shift and rectified linear unit as activation) while following the residual learning approach. The residual function reformulates the layer as learning in correspondence to the layer input. The extracted features from each residual inception block (RIB) are concatenated with corresponding transposed convolutions~\cite{dumoulin2016guide} representing similar spatial dimensions in the expansion phase, using skip connections~\cite{drozdzal2016importance}. Instead of direct concatenation, these skip connections are equipped with spectral spatial and depth attention (SSD) network to process the extracted feature maps and preserve the most relevant high or low-level feature maps. The dimension inconsistency of the concatenating layers is removed using strided convolution to reduce spatial dimensions and $1\times 1$ convolution to reduce the depth, as shown in Fig.~\ref{fig6}. Finally, the sigmoid activated $1\times 1$ convolution outputs the lungs contour map along with segmented regions of COVID-19 abnormalities.
	
	\begin{figure}[!b]
		\centering
		\includegraphics[width=\columnwidth] {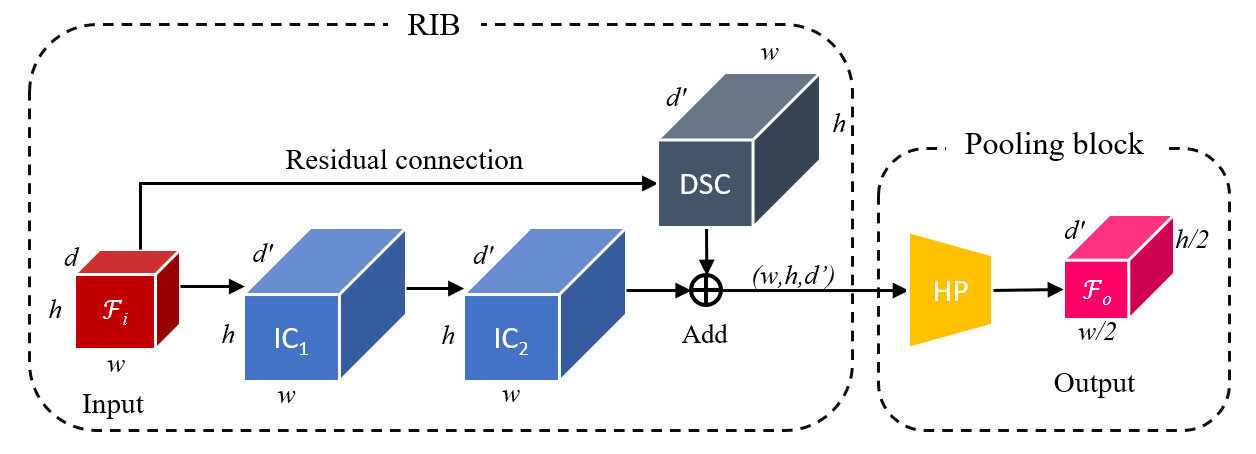}
		\caption{Schematic representation of the residual inception block.}
		\label{fig7}
	\end{figure}
	
	\subsection{Residual inception block}
	Fig.~\ref{fig7} presents the schematic representation of the RIB for some input feature map,  $\mathcal{F}_{i}\in\mathbb{R}^{w\times h\times d}$. It is developed using double inception convolution with a shortcut connection from the input to the output layer. The shortcut or residual connection follows $3\times 3$ DSC whose batch normalized output is merged with the output of the double inception convolution (IC$_1$ and IC$_2$) to extract feature maps using $d’$ number of filters. The consecutive RIBs are connected with the help of valid hybrid pooling which reduces the dimensions of feature maps while preserving the prominent features to produce feature map, $\mathcal{F}_{o}\in\mathbb{R}^{w/2\times h/2\times d'}$.
	
	\subsubsection{Hybrid pooling}
\begin{figure}[H]
		\centering
		\includegraphics[width=0.8\linewidth] {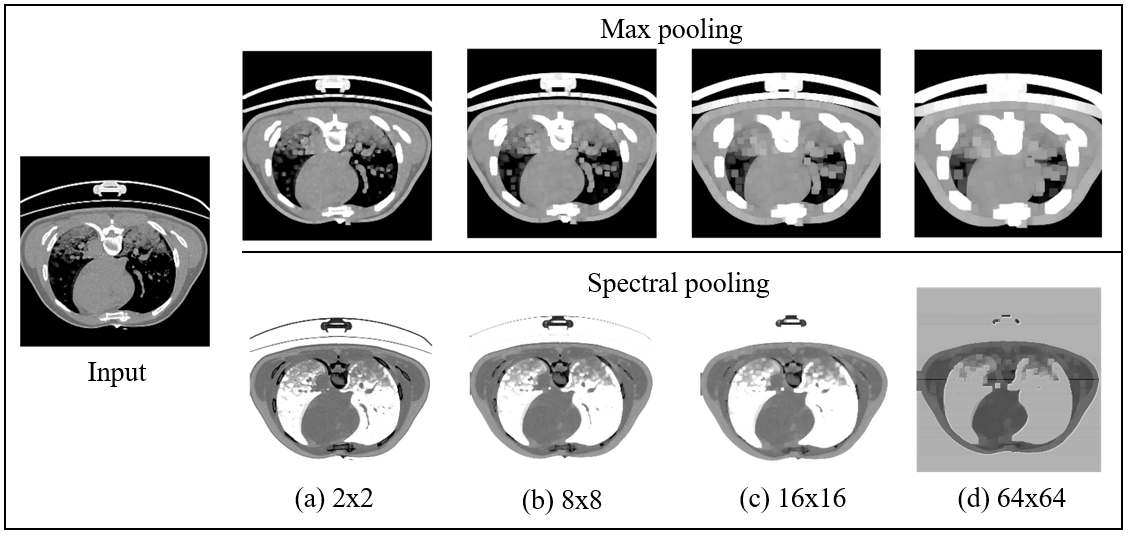}
		\caption{Downsampling using max and spectral pooling with the factors (a) 2$\times$2, (b) 8$\times$8, (c) 16$\times$16 and (d) 64$\times$64.}
		\label{fig8}
	\end{figure}

	Many pooling variants have been proposed based on the value, rank, probability and domain transformation~\cite{akhtar2020interpretation}, among which spectral pooling is found to preserve more spatial information while also reducing the dimensions. Max pooling is featured in every deep learning architecture, however, it only preserves the sharpest features of an image. Fig.~\ref{fig8} presents the comparison of max pooling and spectral pooling for varying filter sizes as $2\times 2$, $8\times 8$, $16\times 16$ and $64\times 64$ on randomly selected CT slice.
	
	In spectral pooling 2D discrete Fourier transform (DFT) of the input image, $\mathcal{I}_{w,h,d}$, is computed as shown in Eq.~\ref {eq14} (concatenated over depth $d$), providing the frequency maps shifted to the center component to truncate the high frequency~\cite{rippel2015spectral}. Finally, the inverse DFT is computed to map the filtered frequency back into the spatial domain, where inverse DFT can be computed as a conjugate of the DFT as ${DFT^{-1}(\cdot)=DFT(\cdot)^*}$.
	
	\begin{equation}
		DFT\left(\mathcal{I}_{w,h,d}\right)_{mn} = \bigcup _{d}\frac{1}{\sqrt{wh}}\sum_{j=0}^{w-1} \sum_{k=0}^{h-1} \mathcal{I}_{jkd} e^{-2\pi i(\frac{jm}{w} + \frac{kn}{h})}
		\label{eq14}
	\end{equation}
	\begin{equation*}
		\forall m\in\{0,\ldots, w-1\}, \forall n\in\{0,\ldots, h-1\}\;
	\end{equation*}
	where $\bigcup _{d}$ indicates concatenation of $DFT$ of each feature map along the channel axis.
	
	In the present article, a hybrid pooling ($\mathcal{P}_h$) approach is proposed to leverage the features of both spectral and max pooling. In this, an input image undergoes spectral and max pooling in parallel followed by $1\times 1$ convolution. This can be performed to generate output with either valid or same padding.
	
	\subsubsection{Inception convolution}
\begin{figure}[!b]
		\centering
		\includegraphics[width=0.8\columnwidth] {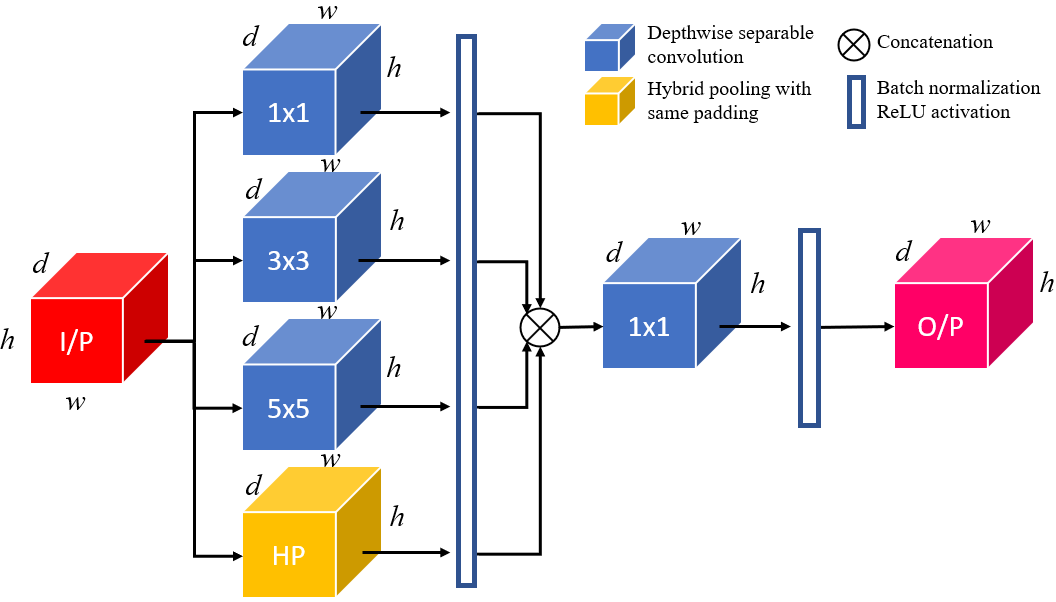}
		\caption{Design of inception convolution (IC).}
		\label{fig9}
	\end{figure}

	It is observed that each feature extraction and reconstruction phase requires different filter sizes to recognize regions or objects of varying dimensions, locale and area. The inception convolution comprises parallel DSCs with filter sizes 1$\times$1, 3$\times$3 and 5$\times$5, and hybrid pooling accompanied with batch normalization (BN) (faster convergence and reduce covariance shift) and rectified linear unit (ReLU) activation function (adds non-linearity) to extract multi-level features for the same instance. The extracted feature maps are then concatenated using 1$\times$1 convolution to optimize the cross channel correlation without modifying the spatial dimensions, followed by BN and ReLU. Fig.~\ref{fig9} describes the schematic representation of the inception convolution that acts as a basis of the RIB.
	
	For an input feature map, $\mathcal{F}_{i}\in\mathbb{R}^{w\times h\times d}$, the inception convolution operation can be represented as in Eq.~\ref{eq15}.
	
	\begin{equation}
		\begin{aligned}
			IC\left(\mathcal{F}_{i}\right)=\left(\left(\left(\bigcup _{f\in \left(1,3,5\right)}\left(\mathcal{F}_i*\mathcal{K}_{f\times r}\right)_{p,q,r}\right)			\bigcup \mathcal{P}_h\left(\mathcal{F}_i\right)\right)*\mathcal{K}_{1\times r}\right)_{p,q,r}
		\end{aligned}
		\label{eq15}
	\end{equation}
	\begin{equation}
		\centering
		\left(\mathcal{F}_{i}*\mathcal{K}_{f\times r}\right)_{p,q,r}=max\left(BN\left(DSC\left( \mathcal{F}_i,\mathcal{K}\right)_{p,q,r}\right),0\right)
	\end{equation}
	where $\bigcup$ indicates concatenation of feature maps along channel axis and $\odot$ indicates element-wise product. $\mathcal{K}_{f\times r}$ represents a kernel or a filter with dimensions $\left(f\times f\times d \times r\right)$ ($r$ indicates number of filters and $d$ indicates same depth of the kernel as input), $\left(\mathcal{F}_i*\mathcal{K}_{f\times r}\right)_{p,q,r}$ indicates a transformed image with features mapping from dimension as $w \times h \times d \mapsto p \times q \times r$ by utilizing the DSC that follows from pointwise convolution operation of the depthwise convolved feature maps. The output dimension, $D_o(p,q,r)$ can be computed as shown in Eq.~\ref{eq16}.
	
	\begin{equation}
		D_{o}\left(p,q,r\right)=\left(\left\lfloor \frac{w+2p-f}{s}+1\right\rfloor ,\left\lfloor \frac{h+2p-f}{s}+1\right\rfloor ,r\right); s>0
		\label{eq16}
	\end{equation}
	where $s$ and $p$ denotes the amount of strides and padding respectively.
	
	\subsection{Spectral spatial and depth attention}
	The attention map aids the network to selectively process the information instead of complete volume by utilizing the inter spatial and channel features correlations. To the best of our knowledge, so far the attention map is obtained by applying either global average pooling or global max pooling, or both~\cite{zagoruyko2016paying, woo2018cbam}. However, these pooling operations tend to be biased towards extreme features (pixels with high intensity). Therefore, a global spectral-max pooling layer is employed in the attention mechanism to extract the most prominent features and generate the attention descriptors, $\mathcal{A}_s^{w\times h\times 1}$ (spatial) and $\mathcal{A}_d^{1\times 1\times d}$ (depth), that infers distinctive object features for an input volume $\mathcal{F}_i$. Finally, these attention descriptors undergo element-wise multiplication with the $\mathcal{F}_i$ followed by ReLU activated $1\times 1$ convolution and BN, to produce refined volumes.
	
	\begin{figure}[!b]
		\centering
		\includegraphics[width=\columnwidth] {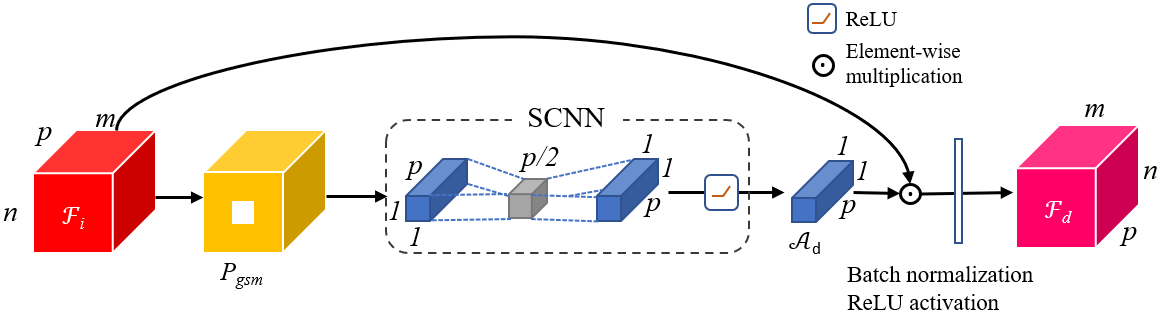}
		\caption{Representation of spectral depth attention network architecture.}
		\label{fig10}
	\end{figure}
	
	\subsubsection{Spectral depth attention}
	The operation of spectral depth attention approach is described in Fig.~\ref{fig10}. The input feature maps, $\mathcal{F}_i\in \mathbb{R}^{m\times n\times p}$ undergoes global spectral-max pooling ($\mathcal{P}_{gsm}$) to generate $\mathcal{F}_{gsm} \in \mathbb{R}^{1\times 1\times p}$. The resulting flattened features pass through the shallow convolution neural network (SCNN). The network comprises of two blocks of $1\times 1$ ReLU activated convolutions to generate depth attention descriptor $\mathcal{A}_d \in \mathbb{R}^{1\times 1\times p}$ as shown in Eq.~\ref{eq17}. The attention descriptor, $\mathcal{A}_d$ then undergoes element-wise multiplication with $\mathcal{F}_i$, followed by batch normalization and ReLU activation to produce spectral depth attention volume $\mathcal{F}_d\in \mathbb{R}^{m\times n\times p}$ as shown in Eq.~\ref{eq18}.
	
	\begin{equation}
		\mathcal{A}_d=\left(\left(SCNN(\mathcal{P}_{gsm}\left(\mathcal{F}_i\right))\right)*\mathcal{K}_{1\times p}\right)_{1,1,p}
		\label{eq17}
	\end{equation}
	\begin{equation}
		\mathcal{F}_d= \left(\left(\mathcal{A}_d\odot \mathcal{F}_i\right)*\mathcal{K}_{1\times p}\right)_{m,n,p}
		\label{eq18}
	\end{equation}
	
	\begin{figure} [!b]
		\centering
		\includegraphics[width=\columnwidth] {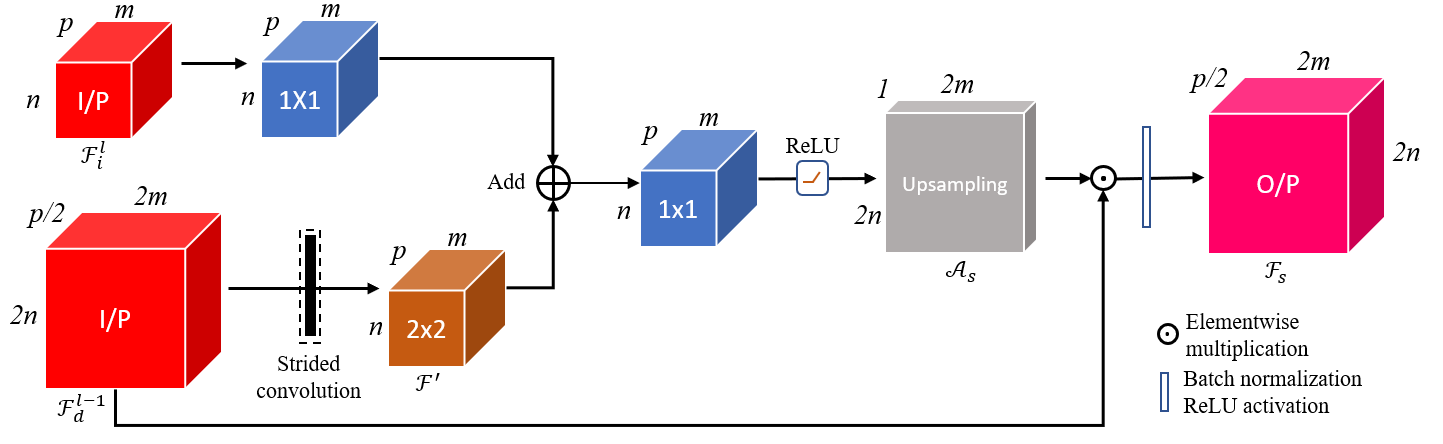}
		\caption{Representation of spectral spatial attention network architecture.}
		\label{fig11}
	\end{figure}
	
	\subsubsection{Spectral spatial attention}
	The overall approach is illustrated in Fig.~\ref{fig11}. For some feature map at layer $l$, $\mathcal{F}_i^{l}\in \mathbb{R}^{m\times n\times p}$, the spectral spatial attention takes the input as $\mathcal{F}_i^{l}$ and spectral depth attention of the previous layer feature map, $\mathcal{F}_d^{l-1}(\mathcal{F}_i^{l-1})$, where  $\mathcal{F}_i^{l-1}\in \mathbb{R}^{2m\times 2n\times p/2}$. The operation starts with the two-strided convolution of $\mathcal{F}_i^{l-1}$ such that it downsamples to $\mathcal{F'}\in \mathbb{R}^{m\times n\times p}$ which is then merged with the $1\times 1$ convolution of $\mathcal{F}_i^{l}$ following the BN and ReLU activation function indicated as $\gamma$ in Eq.~\ref{eq19}. Later, spatial attention descriptor, $\mathcal{A}_s \in \mathbb{R}^{2m\times 2n\times 1}$ is generated by upsampling the ReLU activated $1\times 1$ convolution of the  $\gamma$ as shown in Eq.~\ref{eq20}. Finally, similar to the spectral depth attention, $\mathcal{A}_s$ is element-wise multiplied with $\mathcal{F}_d^{l-1}$ accompanied with $1\times 1$ convolution and batch normalization to form spectral spatial attention volume, $\mathcal{F}_s\in \mathbb{R}^{2m\times 2n\times p/2}$ as shown in Eq.~\ref{eq21}.
	
	\begin{equation}
		\gamma = \left(\mathcal{F}_d^{l-1}*\mathcal{K}_{2\times p}\right)+\left(\mathcal{F}_i^{l}*\mathcal{K}_{1\times p}\right)
		\label{eq19}
	\end{equation}
	
	\begin{equation}
		\mathcal{A}_s = \left(Up\left(\sigma \left(\gamma * \mathcal{K}_{1\times p} \right)\right)\right)_{2m,2n,1}
		\label{eq20}
	\end{equation}
	
	\begin{equation}
		\mathcal{F}_s= \left(\left(\mathcal{A}_s\odot \mathcal{F}_d^{l-1}\right)*\mathcal{K}_{1\times p/2}\right)_{2m,2n,p/2}
		\label{eq21}
	\end{equation}
	\subsection{Objective function}
	The CHS-Net is trained with the segmentation loss function $\mathcal{L}$ defined as the average of binary cross-entropy loss ($\mathcal{L_{BCL}}$) and dice loss ($\mathcal{L_{DL}}$). The segmentation task can be treated as a binary classification task to classify each pixel either belonging to the background (negative) or the region of interest (positive).
	
	The binary cross entropy as defined in Eq.~\ref{eq22} is the most widely used loss function for binary classification and works effectively if there are equal distributions of positive and negative samples.
	
	\begin{equation}
		\begin{aligned}
			\mathcal{L_{BCL}}\left(y,p\left(y\right)\right)=-\sum^N_i\left(y_i.{log \left(p\left(y_i\right)\right)}+\left(1-y_i\right).
			{log \left(1-p\left(y_i\right)\right)}\right)
		\end{aligned}
		\label{eq22}
	\end{equation}
	where $N$ indicates total number of pixels in an image $\mathcal{I}$, $y_i$ and $p(y_i)$ presents the ground truth value and predicted value of $i^{th}$ pixel respectively.
	
	However, due to fewer infectious pixels in CT images, standalone binary cross entropy is not sufficient. Therefore to better penalize the false positive and negative predictions, dice loss is also utilized, defined in Eq.~\ref{eq23}. The $DL$ tends to equally penalize the false negative (FN) and false positive (FP) predictions.
	
	\begin{equation}
		\mathcal{L_{DL}}\left(y,p\left(y\right)\right)=1-\frac{2\sum^N_i{y_i.p(y_i)}}{\sum^N_i{{\left|y_i\right|}^{2}}\mathrm{+}\sum^N_i{{|{p(y}_i)|}^{2}}}
		\label{eq23}
	\end{equation}
	
	The overall loss function $\mathcal{L}$ is represented by Eq.~\ref{eq24}.
	
	\begin{equation}
		\mathcal{L}\left(\mathcal{I},p\left(\mathcal{I}\right)\right)=\frac{1}{2}\mathcal{L_{BCL}}\left(y,p\left(y\right)\right)+\frac{1}{2}\mathcal{L_{DL}}\left(y,p\left(y\right)\right)
		\label{eq24}
	\end{equation}
	
	\section{Experimentation and results}
	\subsection{Dataset description}
	The CHS-Net is trained and evaluated on the synthesized dataset generated using publicly available COVID-19 CT segmentation datasets~\cite{datasetcovid19ct, datasetcovid19ct2}. These two datasets are merged to form an aggregated dataset that addresses the problem of the limited availability of the COVID-19 data. The fused dataset consists of 3560 CT slices with dimensions as $256\times 256\times 1$, each having associated lungs mask and COVID-19 infection mask. Table~\ref{tab1} highlights the class summary details of the fused dataset. 
	
	\begin{table}[!b]
		\centering
		\caption{CT slices distribution details in the synthesized fused dataset.}
		\label{tab1}
		\resizebox{\columnwidth}{!}{\begin{tabular}{llllll}
				\hline
				\multicolumn{1}{l||}{\multirow{4}{*}{Dataset}}                                                                    & \multicolumn{1}{l}{\multirow{4}{*}{\begin{tabular}[c]{@{}l@{}}No. of CT\\ Volumes\end{tabular}}} & \multicolumn{1}{l||}{\multirow{4}{*}{Dimension}}                                      & \multicolumn{3}{l}{Fused dataset}                                                                                                                                                                                                                            \\ \cline{4-6} 
				\multicolumn{1}{l||}{}                                                                                            & \multicolumn{1}{l}{}                                                                             & \multicolumn{1}{l||}{}                                                                & \multicolumn{1}{l}{\begin{tabular}[c]{@{}l@{}}No. of   \\ CT slices\end{tabular}} & \multicolumn{1}{l}{\begin{tabular}[c]{@{}l@{}}No. of\\ Lungs mask\end{tabular}} & \multicolumn{1}{l}{\begin{tabular}[c]{@{}l@{}}No. of\\ COVID-19\\ infection mask\end{tabular}} \\ \hline\hline
				\multicolumn{1}{l||}{\begin{tabular}[c]{@{}l@{}}COVID-19 CT\\ segmentation nr. 2~\cite{datasetcovid19ct}\end{tabular}}              & \multicolumn{1}{l}{9}                                                                            & \multicolumn{1}{l||}{630x630xd*}                                                      & \multicolumn{1}{l}{\multirow{3}{*}{3560}}                                         & \multicolumn{1}{l}{\multirow{3}{*}{3560}}                                       & \multicolumn{1}{l}{\multirow{3}{*}{2200}}                                            \\ \cline{1-3}
				\multicolumn{1}{l||}{\begin{tabular}[c]{@{}l@{}}COVID-19 CT lung and\\ infection segmentation~\cite{datasetcovid19ct2}\end{tabular}} & \multicolumn{1}{l}{20}                                                                           & \multicolumn{1}{l||}{\begin{tabular}[c]{@{}l@{}}630x630xd*\\ 512x512xd*\end{tabular}} & \multicolumn{1}{l}{}                                                              & \multicolumn{1}{l}{}                                                            & \multicolumn{1}{l}{}                                                                 \\ \hline
				\multicolumn{6}{l}{*d indicates that the number of slices varies for   each volume}                                                                                                                                                                                                                                                                                                                                                                                                                                                                                                
		\end{tabular}}
	\end{table}
	\begin{figure}[H]
		\centering
		\includegraphics[scale=0.38] {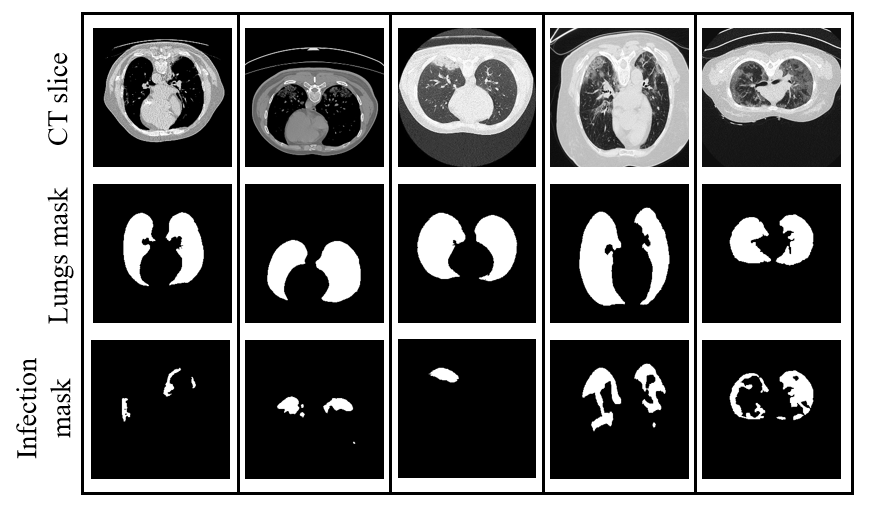}
		\caption{CT slices and corresponding ground truth masks from the fused dataset.}
		\label{fig12}
	\end{figure}
	These 2D CT slices are extracted from the $29$ 3D volumes of the CT imaging having non-uniform or varying dimensions, resized to $256\times 256\times 1$. Each of these slices is annotated carefully by expert radiologists to generate the segmentation mask. Fig.~\ref{fig12} shows the sample slices along with the ground truth segmentation mask corresponding to the lungs and COVID-19 infected region. Each pixel of the slices is marked with class labels as $1$ or $0$ where $1$ means the pixel belongs to the region of interest that is associated with lungs in lungs annotation and COVID-19 (GGO and consolidations) in infection annotation, and $0$ means the background.
	
	\subsection{Training and testing}
\begin{figure}[]
		\centering
		\includegraphics[scale=0.4] {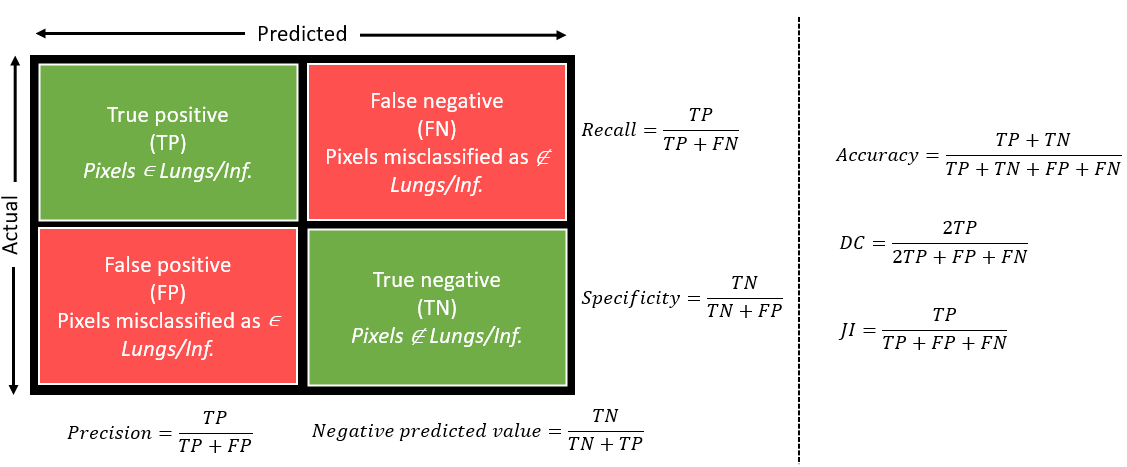}
		\caption{Confusion matrix and evaluation metrices.}
		\label{fig13}
	\end{figure}

	The training and testing sets are acquired from the fused dataset to train and evaluate the proposed CHS-Net model which comprises 70\% and 30\% of the total images respectively. Furthermore, in each set, the distribution of the samples is kept 1:1 from both the datasets~\cite{datasetcovid19ct,datasetcovid19ct2}. The training phase of the CHS-Net is assisted with stochastic gradient descent (SGD) as training weights optimizer to minimize the segmentation loss function (objective function), Adam as a learning rate optimizer~\cite{ruder2016overview}, 5-fold cross-validation for robustness, and early stopping to avoid the overfitting problem~\cite{caruana2001overfitting}. The model is trained in the high performance computing environment with Nvidia RTX Titan GPUs. The trained model is evaluated with the help of the test sets in terms of accuracy, precision, recall, specificity, dice coefficient and Jaccard index (intersection over union). Fig.~\ref{fig13} describes the confusion matrix that can be generated  for the predicted mask corresponding to the CT slice to compute the above discussed metrics based on the correct prediction: true positive (TP), true negative (TN) and incorrect prediction: false positive (FP), false negative (FN). 
	
	Furthermore, uncertainty estimation of the model is performed with Monte Carlo dropout~\cite{gal2016dropout}. Following from work by DeVries et al.~\cite{devries2018leveraging}, the dropout layers are added in the trained model with dropout probability of 0.5 in every convolution layer during testing. From this network the samples are generated 20 times for a given input image and mean segmentation mask ($p$) is acquired. The uncertainty of the model ($U$) for a given input is computed with binary cross entropy function as shown in Eq.~\ref{eq_2} 
	\begin{equation}
		U=-(p_f \log{p_f} + p_b \log{p_b})
		\label{eq_2}
	\end{equation}
	where $p_f$ and $p_b$ is probabilities associated with foreground pixels (target region) and background pixels in the mean segmentation mask respectively.
	
	\begin{figure}[!b]
		\centering
		\includegraphics[width=\linewidth] {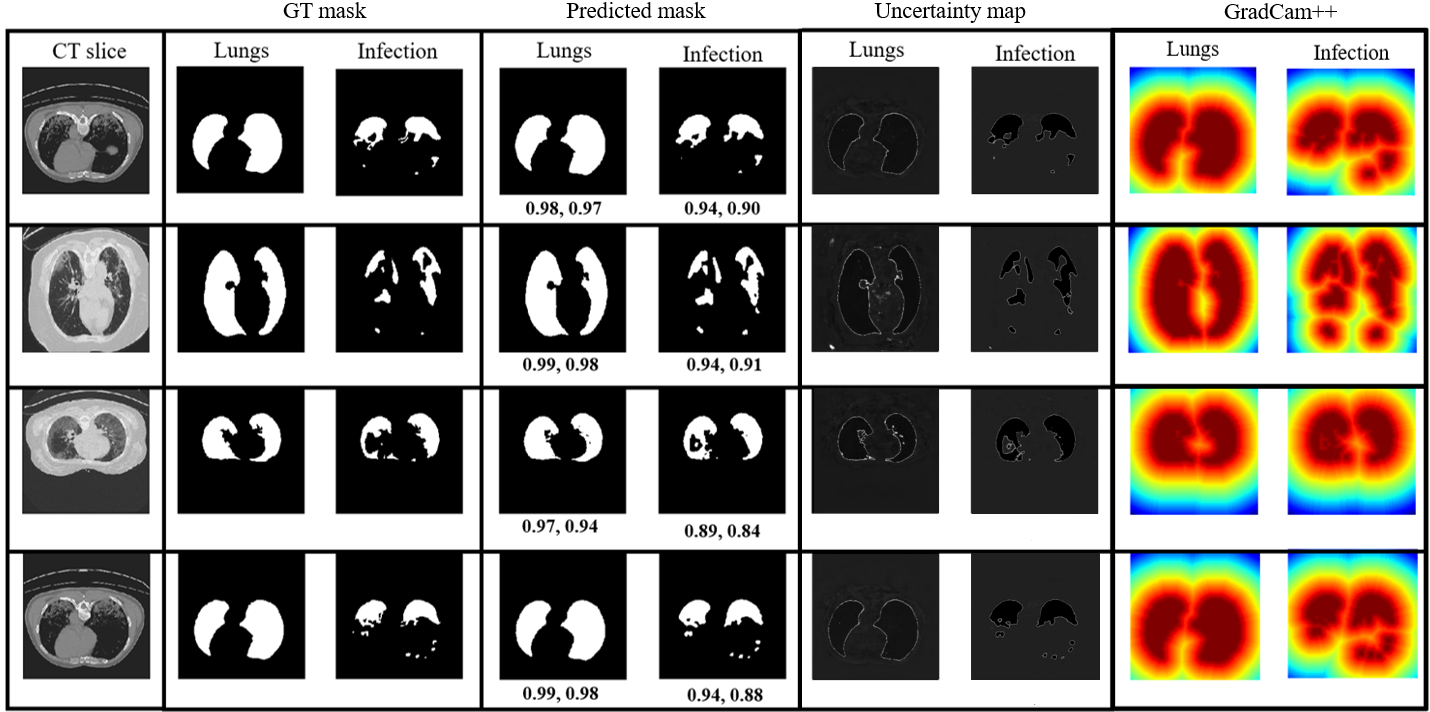}
		\caption{Qualitative results of COVID-19 infection segmentation on test set obtained from fused dataset using CHS-Net. The quantities indicate the dice score and Jaccard index values respectively, for each generated mask.}
		\label{fig14}
	\end{figure}
	
	\subsection{Results and discussion}
	The proposed model generates the semantic segmentation mask of the COVID-19 infected regions in the lungs based on the generated lungs contour maps. Fig.~\ref{fig14} presents the qualitative results of CHS-Net over randomly chosen CT axial slices along with the uncertainty maps, and class activation maps of the output layer using GradCam++~\cite{chattopadhay2018grad} to provide better visual explanations of model predictions. It is observed that the produced segmentation masks match closely with ground truth masks while having tight segmentation boundaries in uncertainty maps, indicating that the model effectively detects and localizes the coronavirus infected regions. Table~\ref{tab2} summarizes the quantitative results of the proposed model for lungs and COVID-19 infection segmentation, where dice coefficients of 0.963 and 0.816 are achieved respectively. 
	
	\begin{table}[]
		\centering
		\caption{Quantitative results of the CHS-Net on the test set generated from fused dataset.}
		\label{tab2}
		\begin{tabular}{l||llllll}
			\hline
			Segmentation & Acc. & Pre. & Spe. & Rec. & DC  & JI   \\ \hline\hline
			Lungs        & 0.991                    & 0.975                    & 0.998                    & 0.950                    & 0.963                  & 0.947                   \\ \hline
			Infection    & 0.965                    & 0.756                    & 0.969                    & 0.885                    & 0.816                  & 0.791                  \\ \hline
		\end{tabular}
	\end{table}
	
	Furthermore, the obtained results are compared with other state-of-the-art segmentation models, as shown in Table~\ref{tab3}. It is observed that CHS-Net approach outperformed the other approaches with significant improvement in the evaluation metrics values, especially in the dice coefficient along with the least number of training parameters. However, among these evaluation metrics, precision is obtained with the least value, indicating that the model generates approximately 25\% of false positive predictions from the pool of test set. However, it is comparatively less than other approaches but leaves the void for further improvements. Furthermore, the models such as attention U-Net~\cite{oktay2018attention}, BCDU-Net~\cite{azad2019bi} and Inf-Net~\cite{fan2020inf} approximately achieved the highest specificity score of 98\% which however is similar to the CHS-Net model in contrast to other models. 
	
	\begin{savenotes}
		\begin{table}[]
			\centering
			\caption{Comparative analysis of the CHS-Net with recently proposed approaches on the fused dataset.}
			\label{tab3}
			\resizebox{\columnwidth}{!}{\begin{tabular}{l||lllllll}
					\hline
					\multicolumn{1}{l||}{\multirow{2}{*}{Model}} & \multicolumn{1}{c||}{\multirow{2}{*}{Param.}} & \multicolumn{6}{c}{Infection segmentation}                                                                                                                                                                        \\ \cline{3-8} 
					\multicolumn{1}{l||}{}                                & \multicolumn{1}{c||}{}                                     & \multicolumn{1}{c}{Acc.} & \multicolumn{1}{c}{Pre.} & \multicolumn{1}{c}{Spe.} & \multicolumn{1}{c}{Rec.} & \multicolumn{1}{c}{DC} & \multicolumn{1}{c}{JI}   \\ \hline\hline
					\multicolumn{1}{l||}{SegNet~\cite{badrinarayanan2017segnet}+VGG16\footnote{\url{https://github.com/lsh1994/keras-segmentation}\label{segnet}}}                   & \multicolumn{1}{l||}{29.4M}                                 & \multicolumn{1}{l}{0.944}             & \multicolumn{1}{l}{0.353}             & \multicolumn{1}{l}{0.791}          & \multicolumn{1}{l}{0.462}          & \multicolumn{1}{l}{0.371}          & \multicolumn{1}{l}{0.320}          \\ \hline
					\multicolumn{1}{l||}{U-Net~\cite{ronneberger2015u}+VGG16\textsuperscript{\ref{segnet}}}                   & \multicolumn{1}{l||}{21.7M}                                 & \multicolumn{1}{l}{0.946}             & \multicolumn{1}{l}{0.383}             & \multicolumn{1}{l}{0.844}          & \multicolumn{1}{l}{0.522}          & \multicolumn{1}{l}{0.441}          & \multicolumn{1}{l}{0.393}          \\ \hline
					\multicolumn{1}{l||}{U-Net++~\cite{zhou2018unet++}+ResNet50\footnote{\url{https://github.com/MrGiovanni/UNetPlusPlus/tree/master/keras}\label{unetpp}}}                   & \multicolumn{1}{l||}{37.6M}                                 & \multicolumn{1}{l}{0.952}             & \multicolumn{1}{l}{0.700}             & \multicolumn{1}{l}{0.903}          & \multicolumn{1}{l}{0.753}          & \multicolumn{1}{l}{0.726}          & \multicolumn{1}{l}{0.674}          \\ \hline
					\multicolumn{1}{l||}{AU-Net~\cite{oktay2018attention}+VGG16\footnote{\url{https://github.com/ozan-oktay/Attention-Gated-Networks}\label{aunet}}}                   & \multicolumn{1}{l||}{22M}                                 & \multicolumn{1}{l}{0.949}             & \multicolumn{1}{l}{0.672}             & \multicolumn{1}{l}{0.978}          & \multicolumn{1}{l}{0.761}          & \multicolumn{1}{l}{0.713}          & \multicolumn{1}{l}{0.664}          \\ \hline
					\multicolumn{1}{l||}{BCDU-Net~\cite{azad2019bi}\footnote{\url{https://github.com/rezazad68/BCDU-Net}\label{bcdunet}}	}                   & \multicolumn{1}{l||}{20.6M}                                  & \multicolumn{1}{l}{0.951}             & \multicolumn{1}{l}{0.705}             & \multicolumn{1}{l}{0.971}         & \multicolumn{1}{l}{0.759}          & \multicolumn{1}{l}{0.731}          & \multicolumn{1}{l}{0.680} \\ \hline
					\multicolumn{1}{l||}{Inf-Net~\cite{fan2020inf}+Res2Net\footnote{\url{https://github.com/DengPingFan/Inf-Net}\label{infnet}}	}                   & \multicolumn{1}{l||}{33M}                                  & \multicolumn{1}{l}{0.954}             & \multicolumn{1}{l}{0.745}             & \multicolumn{1}{l}{\textbf{0.981}}          & \multicolumn{1}{l}{0.764}          & \multicolumn{1}{l}{0.753}          & \multicolumn{1}{l}{0.725}             \\ \hline\hline
					\multicolumn{1}{l||}{RAIU-Net\textsuperscript{\ref{chsnet}}(Ours)}                         & \multicolumn{1}{l||}{4.2M}                                 & \multicolumn{1}{l}{0.950} & \multicolumn{1}{l}{0.712} & \multicolumn{1}{l}{0.961}          & \multicolumn{1}{l}{0.804} & \multicolumn{1}{l}{0.756} & \multicolumn{1}{l}{0.742} \\ \hline
					\multicolumn{1}{l||}{CHS-Net\textsuperscript{\ref{chsnet}}(Ours)}                         & \multicolumn{1}{l||}{8.4M}                                 & \multicolumn{1}{l}{\textbf{0.965}} & \multicolumn{1}{l}{\textbf{0.756}} & \multicolumn{1}{l}{0.969}          & \multicolumn{1}{l}{\textbf{0.885}} & \multicolumn{1}{l}{\textbf{0.816}} & \multicolumn{1}{l}{\textbf{0.791}} \\ \hline
					\multicolumn{8}{l}{*bold quantities indicate highest scores}  
			\end{tabular}}
		\end{table}
	\end{savenotes}
	
	In contrast, the success of the CHS-Net follows from its hierarchical segmentation strategy executed via proposed cascaded RAIU-Net models, where instead of directly segmenting the COVID-19 infected regions, lungs contour maps are generated from the predicted lungs mask which then serve as the input to another RAIU-Net model for localizing the infected regions. Moreover, in CHS-Net, spectral representations (hybrid pooling and global spectral-max pooling) are employed that aids in efficiently downsampling the feature maps with the least loss of information. Furthermore, residual inception blocks tend to efficiently encode and decode the semantic and varying resolution information. In addition, the adopted SSD mechanism refines the high and low extracted feature maps that are later merged in the reconstruction phase. The significance of each component (RIB, HP and SSD) utilized in the proposed framework is presented in Table~\ref{tab4} and Table~\ref{tab5} with the help of the evaluation metrics to highlight their impact on the segmentation performance using the test set. Table~\ref{tab4} illustrates the ablation study for direct COVID-19 infected region segmentation whereas Table~\ref{tab5} shows it for hierarchical segmentation (CHS-Net). This ablation study is carried under the same environment and comprises of baseline U-Net model (BU)~\cite{ronneberger2015u} that follows standard convolutions and max pooling operations. Later, this model is extended with the proposed components as shown in Table~\ref{tab4} and Table~\ref{tab5} to highlight the significance of each component for achieving the concerned results. As observed in both the tables, even for the baseline model the specificity metric value is significantly higher as compared to other metrics, indicating that model is predicting all pixels as background (dark) and hence not able to detect any infection, whereas with continuous incorporation of the proposed components the model tends to perform better in localizing the complex patterns associated with the infection. For instance, in Table~\ref{tab5}, the dice coefficient for lungs and infection segmentation by BU mode improves by 13\% and 57\% with the addition of RIB (BU+RIB) respectively. In addition, it is also observed that the hierarchical segmentation approach works significantly better than direct segmentation, where each component certainly contributes to improvement in the model performance.
	
	\begin{table}[]
		\centering
		\caption{Effects of the proposed components on the model performance for direct segmentation of COVID-19 infectious regions.}
		\label{tab4}
		\begin{tabular}{l||llllll}
			\hline
			\multirow{2}{*}{Mode}    & \multicolumn{6}{c}{Infection segmentation}                                                                                                                   \\\cline{2-7}
			& \multicolumn{1}{l}{Acc.} & \multicolumn{1}{l}{Pre.} & \multicolumn{1}{l}{Spe.} & \multicolumn{1}{l}{Rec.} & \multicolumn{1}{l}{DC} & \multicolumn{1}{l}{JI}  \\ \hline \hline
			BU                       & 0.680                     & 0.043                    & 0.803                    & 0.281                    & 0.075                  & 0.031                   \\ \hline
			BU+RIB                   & 0.751                    & 0.551                    & 0.901                    & 0.354                    & 0.432                  & 0.401                   \\ \hline
			BU+HP                    & 0.713                    & 0.138                    & 0.834                    & 0.295                    & 0.189                  & 0.122                   \\ \hline
			BU+RIB+HP                & 0.880                     & 0.636                    & 0.951                    & 0.688                    & 0.661                  & 0.631                   \\ \hline
			BU+HP+SSD                & 0.825                    & 0.510                     & 0.935                    & 0.556                    & 0.533                  & 0.523                   \\ \hline
			BU+RIB+SSD               & 0.924                    & 0.635                    & 0.953                    & 0.783                    & 0.702                  & 0.689                   \\ \hline
			BU+RIB+HP+SSD (RAIU-Net) & 0.950                     & 0.712                    & 0.961                    & 0.804                    & 0.756                  & 0.742                  \\ \hline
		\end{tabular}
	\end{table}
	
	\begin{table}[]
		\centering
		\caption{Effects of the proposed components on the model performance for hierarchical segmentation of COVID-19 infectious regions.}
		\label{tab5}
		\resizebox{\columnwidth}{!}{\begin{tabular}{l||l||llllll}
				\hline
				Mode                                                                                & Segmentation & \multicolumn{1}{l}{Acc.} & \multicolumn{1}{l}{Pre.} & \multicolumn{1}{l}{Spe.} & \multicolumn{1}{l}{Rec.} & \multicolumn{1}{l}{DC} & \multicolumn{1}{l}{JI}  \\ \hline \hline
				\multirow{2}{*}{BU}                                                                 & Lungs        & 0.921                    & 0.731                    & 0.951                    & 0.780                    & 0.755                  & 0.731                   \\ \cline{2-8}
				& Infection    & 0.753                    & 0.212                    & 0.799                    & 0.321                    & 0.256                  & 0.221                   \\ \hline
				\multirow{2}{*}{BU+RIB}                                                             & Lungs        & 0.984                    & 0.853                    & 0.981                    & 0.884                    & 0.869                  & 0.841                   \\ \cline{2-8}
				& Infection    & 0.879                    & 0.687                    & 0.944                    & 0.533                    & 0.601                  & 0.589                   \\ \hline
				\multirow{2}{*}{BU+HP}                                                              & Lungs        & 0.948                    & 0.758                    & 0.963                    & 0.791                    & 0.775                  & 0.741                   \\ \cline{2-8}
				& Infection    & 0.781                    & 0.215                    & 0.811                    & 0.336                    & 0.263                  & 0.236                   \\ \hline
				\multirow{2}{*}{BU+RIB+HP}                                                          & Lungs        & 0.980                    & 0.915                    & 0.991                    & 0.927                    & 0.921                  & 0.901                   \\ \cline{2-8}
				& Infection    & 0.954                    & 0.731                    & 0.956                    & 0.820                    & 0.773                  & 0.754                   \\ \hline
				\multirow{2}{*}{BU+HP+SSD}                                                          & Lungs        & 0.955                    & 0.806                    & 0.980                    & 0.851                    & 0.828                  & 0.791                   \\ \cline{2-8}
				& Infection    & 0.901                    & 0.588                    & 0.910                    & 0.601                    & 0.595                  & 0.633                   \\ \hline
				\multirow{2}{*}{BU+RIB+SSD}                                                         & Lungs        & 0.964                    & 0.948                    & 0.990                    & 0.940                    & 0.944                  & 0.932                   \\ \cline{2-8}
				& Infection    & 0.937                    & 0.753                    & 0.960                    & 0.870                    & 0.808                  & 0.788                   \\ \hline
				\multirow{2}{*}{\begin{tabular}[c]{@{}l@{}}BU+RIB+HP+SSD \\(CHS-Net)~\end{tabular}} & Lungs        & 0.991                    & 0.975                    & 0.998                    & 0.950                    & 0.963                  & 0.947                   \\ \cline{2-8}
				& Infection    & 0.965                    & 0.756                    & 0.969                    & 0.885                    & 0.816                  & 0.791                  \\ \hline
		\end{tabular}}
	\end{table}

	\section{Conclusion}
	This article proposes a COVID-19 hierarchical segmentation network, CHS-Net, to identify the COVID-19 infected regions from the generated lungs contour maps with computed tomography (CT) images via cascaded residual attention inception U-Net (RAIU-Net) models. The RAIU-Net model improves upon the base U-Net model by exploiting various state-of-the-art components to improve the feature extraction and reconstruction process, where a residual inception block (RIB) and spectral spatial and depth attention (SSD) network tends to effectively encode and decode the feature maps at varying resolutions, while also addressing the major challenges involved in the segmentation. With extensive trials it is observed that the results of the CHS-Net model outperformed the other recently proposed approaches which were evaluated using standard benchmark performance metrics i.e. accuracy (96\%), precision (75\%), recall (88\%), specificity (97\%), dice coefficient (81\%) and Jaccard index (intersection over union) (89\%). The ablation study of the CHS-Net highlighted the contribution of each component towards the improvement in the segmentation results. Besides, trials can be made to tweak and tune the architecture with other deep learning components to further improve the results. It is believed that the potential of CHS-Net design can also be extended to other applications concerning biomedical image segmentation.
	
	\section*{Acknowledgment}
	We thank our institute, Indian Institute of Information Technology Allahabad (IIITA), India and Big Data Analytics (BDA) lab for allocating the centralised computing facility and other necessary resources to perform this research. We extend our thanks to our colleagues for their valuable guidance and suggestions.
	
	\bibliographystyle{elsarticle-num}
	\bibliography{reference}
	
\end{document}